\documentclass[prd,twocolumn,showpacs,amsmath,amssymb]{revtex4}
\usepackage{graphicx}% Include figure files
\usepackage{dcolumn}% Align table columns on decimal point
\usepackage{bm}% bold math
\newcommand{\pip}{\pi^+}
\newcommand{\pim}{\pi^-}
\newcommand{\piz}{\pi^0}
\newcommand{\kap}{K^+}
\newcommand{\kam}{K^-}
\newcommand{\kaz}{K^0}
\newcommand{\kazb}{\overline{K}^0}

\newcommand{\kstz}{K^{*0}}
\newcommand{\kstzb}{\overline{K}^{*0}}

\newcommand{\roz}{\rho^0}

\newcommand{\etap}{\eta^{\prime}}
\newcommand{\etaz}{\eta_{0}}

\newcommand{\psp}{\psi^{\prime}}

\newcommand{\jpsi}{J/\psi}

\newcommand{\EE}{e^+e^-}

\newcommand{\nnb}{n\overline{n}}

\newcommand{\ppb}{p\overline{p}}

\newcommand{\LLb}{\Lambda \overline{\Lambda}}

\newcommand{\SsSsbz}{\Sigma^{*0} \overline{\Sigma^*}^0}
\newcommand{\SsSsbp}{\Sigma^{*+} \overline{\Sigma^*}^-}
\newcommand{\SsSsbm}{\Sigma^{*-} \overline{\Sigma^*}^+}
\newcommand{\XsXsbz}{\Xi^{*0} \overline{\Xi^*}^0}
\newcommand{\XsXsbm}{\Xi^{*-} \overline{\Xi^*}^+}
\newcommand{\SSbz}{\Sigma^0 \overline{\Sigma}^0}
\newcommand{\SSbp}{\Sigma^+ \overline{\Sigma}^-}
\newcommand{\SSbm}{\Sigma^- \overline{\Sigma}^+}
\newcommand{\XXbz}{\Xi^0 \overline{\Xi}^0}
\newcommand{\XXbm}{\Xi^- \overline{\Xi}^+}
\newcommand{\SzLb}{\Sigma^0 \overline{\Lambda}}

\newcommand{\SbzL}{\overline{\Sigma}^0 \Lambda}
\newcommand{\DDltpp}{\Delta^{++} \overline{\Delta}^{--}}
\newcommand{\DDltp}{\Delta^{+} \overline{\Delta}^{-}}
\newcommand{\DDltz}{\Delta^{0} \overline{\Delta}^{0}}
\newcommand{\DDltn}{\Delta^{-} \overline{\Delta}^{+}}
\newcommand{\OOb}{\Omega^{-} \overline{\Omega}^{+}}
\newcommand{\SbsmSp}{\overline{\Sigma^*}^- \Sigma^+}
\newcommand{\SspSbm}{\Sigma^{*+} \overline{\Sigma}^-}
\newcommand{\SbszSz}{\overline{\Sigma^*}^0 \Sigma^0}
\newcommand{\SszSbz}{\Sigma^{*0} \overline{\Sigma}^0}
\newcommand{\SbspSm}{\overline{\Sigma^*}^+ \Sigma^-}
\newcommand{\SsmSbp}{\Sigma^{*-} \overline{\Sigma}^+}
\newcommand{\XbszXz}{\overline{\Xi^*}^0 \Xi^0}
\newcommand{\XszXbz}{\Xi^{*0} \overline{\Xi}^0}
\newcommand{\XbspXm}{\overline{\Xi^*}^+ \Xi^-}
\newcommand{\XsmXbp}{\Xi^{*-} \overline{\Xi}^+}
\newcommand{\Dbmpp}{\overline{\Delta}^- p}
\newcommand{\Dppbm}{\Delta^+ \overline{p}}
\newcommand{\Dbznz}{\overline{\Delta}^0 n}
\newcommand{\Dznbz}{\Delta^0 \overline{n}}
\newcommand{\SbszL}{\overline{\Sigma^*}^0 \Lambda}
\newcommand{\SszLb}{\Sigma^{*0} \overline{\Lambda}}
\newcommand{\xet}{X_{\eta}}
\newcommand{\xetp}{X_{\eta^{\prime}}}
\newcommand{\xetz}{X_{\eta_{0}}}
\newcommand{\yet}{Y_{\eta}}
\newcommand{\yetp}{Y_{\eta^{\prime}}}
\newcommand{\yetz}{Y_{\eta_{0}}}
\newcommand{\zet}{Z_{\eta}}
\newcommand{\zetp}{Z_{\eta^{\prime}}}
\newcommand{\zetz}{Z_{\eta_{0}}}
\newcommand{\Heff}{{\cal H}_{eff}}

\newcommand{\gz}{g_{0}}
\newcommand{\beq}{\begin{equation}}
\newcommand{\eeq}{\end{equation}}
\def\eref#1{(\ref{#1})}
\def\Journal#1#2#3#4{{#1} {\bf #2}, #3 (#4)}
\def\PRD{Phys. Rev. D}
\begin{document}

\title{Generic symmetry analysis of charmonium decay}

\author{X.H.Mo$^{1,2}$
\\  \vspace{0.2cm} {\it
$^{1}$ Institute of High Energy Physics, CAS, Beijing 100049, China\\
$^{2}$ University of Chinese Academy of Sciences, Beijing 100049, China\\
}
}
%\affiliation{Institute of High Energy Physics, CAS, Beijing 100049, China}
\email{moxh@ihep.ac.cn}
\date{\today}
	
\begin{abstract}
For charmonium's decaying to the final states involving merely light quarks, in light of $SU(3)$ flavor symmetry, a systematic parametrization scheme is established, which involving binary decays, ternary decays and radiative decays.
\end{abstract}
% PACS
\maketitle

%\section{Introduction}
Quantum chromodynamics (QCD) as a widely appreciated theory of strong interaction, has been proved to be very successful at high energy when the calculation can be executed perturbatively. Nevertheless, its validity at non-perturbative regime, such as $\jpsi$ and $\psp$ resonance regions, needs more experimental guidance. The production and decay of charmonium states supply an ideal laboratory for such a study.

As a matter of fact, many models are constructed for charmonium decay~\cite{Kowalski:1976mc}-\cite{moxh2024}, the parametrization of various tow body decay modes are obtained, especially a systematic parametrization scheme is proposed recently in Refs.~\cite{moxh2023,moxh2024}. By virtue of $SU(3)$ flavor symmetry, the effective interaction Hamiltonian is obtained according to group representation theory.

In this Letter, a systematic and exclusive parametrization scheme is established for all kinds of charmonium decay. First, two improvements are made for two-body decays. One is the addition of a new kind of breaking effect, that is the effect due to quark magnetic momentum. The other is the extension of the mixing scenario to include the admixture between glueball-like scalar and pseudoscalar. Second, the parametrization framework for three-body decay is obtained. Third, the symmetry analysis extends to the radiative decay.

%\section{Analysis framework}\label{xct_alsfrk}
We know that in the $\EE$ collider experiment, the initial state is obviously flavorless, then the final state must be flavor singlet. Moreover, only the Okubo-Zweig-Iizuka (OZI) rule suppressed processes are considered, and the final states merely involve light quarks, that is $u, d, s$ quarks. Therefore, $SU(3)$ group is employed for symmetry analysis. The key rule herein is the so-call ``flavor singlet principle'' that determines what kinds of terms are permitted in the effective interaction Hamiltonian. Resorting to the perturbation language, the Hamiltonian is written as
\beq
\Heff = H_0 + \Delta H~,
\label{perturbaionhmtn}
\eeq
where $H_0$ is the symmetry conserved term and $\Delta H$ the symmetry breaking term, which is generally small compare to $H_0$. In the light of group representation theory, the product of two multiplets, say ${\mathbf n}$ and ${\mathbf m}$, can be decomposed into a series of irreducible representations, that is
\beq
{\mathbf n} \otimes {\mathbf m} = {\mathbf l_1} \oplus {\mathbf l_2} \oplus \cdots \oplus {\mathbf
l_k}~.
\label{dcpsmoftwomtplt}
\eeq
The singlet principle requires that among the ${\mathbf l_j} (j=1, \cdots, k)$, only the singlet term, i.e., ${\mathbf l_j}={\mathbf 1}$ for certain $j$, is allowed in the Hamiltonian. Since this term is obviously $SU(3)$ invariant, it is called the symmetry conserved term, i.e., $H_0$.

\begin{table*}[bth]
\caption{\label{bynamppmznfm} Amplitude parametrization forms for decays of a resonance into octet-octet ($O-O$) mode, decuplet-decuplet ($D-D$) mode, and decuplet-octet ($D-O$) mode. Symbols $A$, $D$, and $F$ are introduced for simplifying the expression, the relation of them with effective coupling constants $g_m$, $g_e$, and $g_{\mu}$ are indicated in this table. }
%\center
\begin{ruledtabular}
\begin{tabular}{lccccccc|lcccc|lccc}
\multicolumn{8}{c|}{$O-O$ mode} & \multicolumn{5}{c|}{$D-D$ mode} & \multicolumn{4}{c}{$D-O$ mode}\\ \hline
  Final   &$A$ &$D_m$&$F_m$&$D_e$&$F_e$&$D_{\mu}$&$F_{\mu}$& Final   &$A$ &$D_m$&$D_e$&$D_{\mu}$ & Final   &$D_m$&$D_e$&$D_{\mu}$\\  \cline{2-8} \cline{10-13} \cline{15-17}
  state   &$g_0$ &$g_m^{\prime}/3$&$g_m$&$g_e^{\prime}/3$&$g_e$&$g_{\mu}^{\prime}/3$&$g_{\mu}$
  & state   &$g_0$ &$g_m/3$&$g_e/3$&$g_{\mu}/3$
  & state   &$g_m/\sqrt{3}$&$g_e/\sqrt{3}$&$g_{\mu}/\sqrt{3}$\\ \hline
  $\ppb$  &$1$ &$1$  &$1$  &$1$  &$-1$ &$-2$     &$0$      &$\DDltpp$&$1$ &$-1$ &$2$  &$-1$  &$\SbsmSp /\SspSbm$  &$1$  &$-1$   &$0$   \\
  $\nnb$  &$1$ &$1$  &$1$  &$-2$ &$0$  &$1$      &$-1$     &$\DDltp$ &$1$ &$-1$ &$1$  &$0$   &$\SbszSz /\SszSbz$  &$-1$ &$1/2$ &$1/2$ \\
  $\SSbp$ &$1$ &$-2$ &$0$  &$1$  &$-1$ &$1$      &$1$      &$\DDltz$ &$1$ &$-1$ &$0$  &$1$   &$\SbspSm /\SsmSbp$  &$-1$ &$0$   &$1$ \\
  $\SSbz$ &$1$ &$-2$ &$0$  &$1$  &$0$  &$1$      &$0$      &$\DDltn$ &$1$ &$-1$ &$-1$ &$2$   &$\XbszXz /\XszXbz$  &$1$ &$-1$  &$0$ \\
  $\SSbm$ &$1$ &$-2$ &$0$  &$1$  &$1$  &$1$      &$-1$     &$\SsSsbp$  &$1$ &$0$  &$1$  &$-1$  &$\XbspXm /\XsmXbp$  &$-1$&$0$  &$1$ \\
  $\XXbz$ &$1$ &$1$  &$-1$ &$-2$ &$0$  &$1$      &$1$      &$\SsSsbz$  &$1$ &$0$  &$0$  &$0$   &$\Dbmpp /\Dppbm$    &$0$ &$1$  &$-1$ \\
  $\XXbm$ &$1$ &$1$  &$-1$ &$1$  &$1$  &$-2$     &$0$      &$\SsSsbm$  &$1$ &$0$  &$-1$ &$1$   &$\Dbznz /\Dznbz$     &$0$&$1$  &$-1$ \\
  $\LLb$  &$1$ &$2$  &$0$  &$-1$ &$0$  &$-1$     &$0$      &$\XsXsbz$  &$1$ &$1$  &$0$  &$-1$  &$\SbszL /\SszLb$    &$0$ &$-\sqrt{3}/2$   &$\sqrt{3}/2$\\
  $\SzLb$ &$0$ &$0$  &$0$ &$\sqrt{3}$&$0$ &$-\sqrt{3}$&$0$ &$\XsXsbm$  &$1$ &$1$  &$-1$ &$0$   & & & &    \\
  $\SbzL$ &$0$ &$0$  &$0$ &$\sqrt{3}$&$0$ &$-\sqrt{3}$&$0$ &$\OOb$   &$1$ &$2$  &$-1$ &$-1$  & & & &    \\
\end{tabular}
\end{ruledtabular}
\end{table*}

As far as the $SU(3)$-breaking effect is concerned, they are treated as a ``spurion'' octet, then the favor singlet principle is used to pin down the breaking term in Hamiltonian. There are totally three kinds of effects, that is the strong breaking effect, the electromagnetic breaking effect, and the breaking effect due to the magnetic momentum of quarks, which can be expressed as
\beq
\left.\begin{array}{rl}
S_m = & diag [1,1, -2]~, \\
S_e = & diag [2, -1,-1]~,  \\
S_{\mu} =& diag [2, -1,2]~.
\end{array}\right.~~
\label{brktms}
\eeq
It is worthy of noticing that these three breaking effects de facto fully exhaust the possible symmetry of elementary representation. From pure viewpoint of group theory, they are not independent, but the physical original of them is obviously distinct.

Now we first consider baryon parametrization. Following the deductions of Ref.~\cite{moxh2023}, the parametrization results are tabulated in Table~\ref{bynamppmznfm}. The forms are similar to those of Ref.~\cite{moxh2023} and the only difference is the addition of the contribution from the $SU(3)$-breaking effect due to the magnetic momentum of quarks.

Second, we consider meson parametrization. Here the generalized inherent ${C}$-parity for a multiplet is introduced, and its value is set to be equal to that of the neutral particle in the multiplet. For two octet meson final states, denoted respectively by $O_1$ and $O_2$, defined are the following terms, which may be allowed or forbidden in the effective Hamiltonian:
\beq
[O_1 O_2]_0 = (O_1)^i_j (O_2)^j_i~~,
\eeq
\beq
([O_1 O_2]_f )^i_j = (O_1)^i_k (O_2)^k_j -(O_1)^k_j (O_2)^i_k~~,
\eeq
and
\beq
([O_1 O_2]_d )^i_j = (O_1)^i_k (O_2)^k_j +(O_1)^k_j (O_2)^i_k
-\frac{2}{3} \delta^i_j \cdot (O_1)^i_j (O_2)^j_i~~.
\eeq
Under parity transformation, $\hat{C} [O_1 O_2]_{x} \to \xi_{x} [O_1 O_2]_{x}$, where $x=0,d,f$, that is $\xi_{0}=+1,\xi_{d}=+1,\xi_{f}=-1$. In addition, $\hat{C} O_i \to \eta_{O_i} O_i, (i=1,2)$, synthetically,
\beq
\hat{C}~[O_1 O_2]_{x}  = \eta_{O_1} \eta_{O_2} \xi_x [O_1 O_2]_{x}~,  \\
\label{ctfmnforokt}
\eeq
At the same time for the initial state of $\psi$, $\hat{C}~\psi  = \eta_{\psi} \psi$. Then the term $[O_1 O_2]_{x}$ is allowed in the effective Hamiltonian as long as $\eta_{\psi}=-1=\eta_{O_1} \eta_{O_2} \xi_x$, otherwise, it is forbidden. With this criterion, it is easy to figure out what kind of terms can be adopted in the effective Hamiltonian for various kinds of final states. As a matter of fact, there exist merely two types of Hamiltonian forms. One contains both $[O_1 O_2]_{0}$ and $[O_1 O_2]_{d}$ terms, while the other contains only $[O_1 O_2]_{f}$ term, that is
\beq
\left.\begin{array}{rl}
\Heff^{O_1 O_2}= & \gz \cdot [O_1 O_2]_0 + g_m \cdot ([O_1 O_2]_d )^3_3 \\
        & + g_e \cdot ([O_1 O_2]_d )^1_1 + g_{\mu} \cdot ([O_1 O_2]_d )^2_2 ~,
\end{array}\right.~~
\label{effhmtvptype}
\eeq
or
\beq
\Heff^{O_1 O_2} =  g_m \cdot ([O_1 O_2]_f )^3_3  + g_e \cdot ([O_1 O_2]_f )^1_1
 + g_{\mu} \cdot ([O_1 O_2]_f )^2_2 ~.
\label{effhmtpptype}
\eeq
If $O_1=V$ (Vector) and $O_2=P$ (Pseduoscalar), the corresponding parametrization can be obtained and presented in Table~\ref{vpmsnform}.

\begin{table}[hbt]
\caption{\label{vpmsnform} Amplitude parametrization form for
decays of the $\psp$ or $\jpsi$ into $V P$ final states. General expressions in terms of singlet $A$ (by definition $A=g_0$), as well as the mass-breaking term ($D_m=g_m/3$), the charge-breaking term ($D_e=g_e/3$), and  the term due to magnetic momentum ($D_{\mu}=g_{\mu}/3$). } \center
\begin{tabular}{lc}\hline \hline
  Final state    & Amplitude parametrization form  \\ \hline
  $\rho^{\pm}\pi^{\mp}$, $\roz \piz$  & $A-2D_m+D_e+D_{\mu}$    \\
  $K^{*\pm} K^{\mp}$                  & $A+D_m+D_e-2D_{\mu}$     \\
  $\kstz\kazb$, $\kstzb \kaz$         & $A+D_m-2D_e+D_{\mu}$    \\
  $\omega \eta$                       & $A+2D_m-D_e-D_{\mu}$    \\
  $\omega \piz$                       & $\sqrt{3}D_e-\sqrt{3}D_{\mu}$     \\
  $\roz \eta$                         & $\sqrt{3}D_e-\sqrt{3}D_{\mu}$    \\
\hline \hline
\end{tabular}
\end{table}

\begin{table*}[bth]
\caption{\label{vpmsnfmmix}Amplitude parametrization form for decays of the $\psp$ or $\jpsi$ into $V~P$ final states. The shorthand symbols are defined as $s_{\alpha}\equiv \sin \theta_{\alpha}$, $c_{\alpha}\equiv \cos \theta_{\alpha}$, $s^{\pm}_{\alpha\beta}\equiv \sin ( \theta_{\alpha}\pm \theta_{\beta})$,$c^{\pm}_{\alpha\beta}\equiv \cos ( \theta_{\alpha}\pm \theta_{\beta})$, $s_{\gamma}\equiv \sin \theta_{\gamma}=\sqrt{1/3}$, and $c_{\gamma}\equiv \cos \theta_{\gamma}=\sqrt{2/3}$.}
\begin{ruledtabular}
\begin{tabular}{ccccc}
Decay mode    &\multicolumn{4}{c}{Coupling constant} \\  \cline{2-5}
$\psi \to X$                       &$g_0$   &$g_m$     &$g_e$   &$g_{\mu}$ \\ \hline
$\rho^{\pm}\pi^{\mp}$, $\roz \piz$ &  1     &$-2/3$    &$1/3$   &$1/3$   \\
$K^{*\pm} K^{\mp}$                 &  1     &$1/3$     &$1/3$   &$-2/3$  \\
$\kstz\kazb$, $\kstzb \kaz$        &  1     &$1/3$     &$-2/3$  &$1/3$   \\
$\phi\eta$  &$s^{-}_{\gamma V}\xet-c^{-}_{\gamma V}\yet$
            &$-\frac{2}{3} (s^{-}_{\gamma V}\xet+2 c^{-}_{\gamma V}\yet)$
            &$\frac{1}{3} (s^{-}_{\gamma V}\xet+2 c^{-}_{\gamma V}\yet)$
            &$\frac{1}{3} (s^{-}_{\gamma V}\xet+2 c^{-}_{\gamma V}\yet)$  \\
$\phi\etap$ &$s^{-}_{\gamma V}\xetp-c^{-}_{\gamma V}\yetp$
            &$-\frac{2}{3} (s^{-}_{\gamma V}\xetp+2 c^{-}_{\gamma V}\yetp)$
            &$\frac{1}{3} (s^{-}_{\gamma V}\xetp+2 c^{-}_{\gamma V}\yetp)$
            &$\frac{1}{3} (s^{-}_{\gamma V}\xetp+2 c^{-}_{\gamma V}\yetp)$  \\
$\phi\etaz$ &$s^{-}_{\gamma V}\xetz-c^{-}_{\gamma V}\yetz$
            &$-\frac{2}{3} (s^{-}_{\gamma V}\xetz+2 c^{-}_{\gamma V}\yetz)$
            &$\frac{1}{3} (s^{-}_{\gamma V}\xetz+2 c^{-}_{\gamma V}\yetz)$
            &$\frac{1}{3} (s^{-}_{\gamma V}\xetz+2 c^{-}_{\gamma V}\yetz)$  \\
$\omega\eta$&$c^{-}_{\gamma V}\xet+s^{-}_{\gamma V}\yet$
            &$-\frac{2}{3} (c^{-}_{\gamma V}\xet+2 s^{-}_{\gamma V}\yet)$
            &$\frac{1}{3} (c^{-}_{\gamma V}\xet+2 s^{-}_{\gamma V}\yet)$
            &$\frac{1}{3} (c^{-}_{\gamma V}\xet+2 s^{-}_{\gamma V}\yet)$  \\
$\omega\etap$&$c^{-}_{\gamma V}\xetp+s^{-}_{\gamma V}\yetp$
            &$-\frac{2}{3} (c^{-}_{\gamma V}\xetp+2 s^{-}_{\gamma V}\yetp)$
            &$\frac{1}{3} (c^{-}_{\gamma V}\xetp+2 s^{-}_{\gamma V}\yetp)$
            &$\frac{1}{3} (c^{-}_{\gamma V}\xetp+2 s^{-}_{\gamma V}\yetp)$  \\
$\omega\etaz$&$c^{-}_{\gamma V}\xetz+s^{-}_{\gamma V}\yetz$
            &$-\frac{2}{3} (c^{-}_{\gamma V}\xetz+2 s^{-}_{\gamma V}\yetz)$
            &$\frac{1}{3} (c^{-}_{\gamma V}\xetz+2 s^{-}_{\gamma V}\yetz)$
            &$\frac{1}{3} (c^{-}_{\gamma V}\xetz+2 s^{-}_{\gamma V}\yetz)$  \\
$\roz\eta$  &   0 &   0 & $\xet$  &$-\xet$ \\
$\roz\etap$ &   0 &   0 & $\xetp$ &$-\xetp$ \\
$\roz\etaz$ &   0 &   0 & $\xetz$ &$-\xetz$ \\
$\phi\piz$  &   0 &   0 & $s^{-}_{\gamma V}$ & $-s^{-}_{\gamma V}$ \\
$\omega\piz$&   0 &   0 & $c^{-}_{\gamma V}$ & $-c^{-}_{\gamma V}$ \\
\end{tabular}
\end{ruledtabular}
\end{table*}

In above analysis, mesons are treated as pure octet components, but the observed ones are actually the mixing of pure octet and singlet components. Moreover, whatever theoretically or experimentally one can not exclude the possible admixture of quarkonium with gluonium states. In a more general mixing framework,
\beq
\left(\begin{array}{c}
\eta  \\
\etap \\
\etaz
\end{array}\right) = {\mathbf O_R}
\left(\begin{array}{c}
\eta^8  \\
\eta^1 \\
G
\end{array}\right)~.
\label{mixeepez}
\eeq
As an element of the orthogonal group $O(3)$, ${\mathbf O_R}$ depends on three mixing angles $\theta_1$, $\theta_2$, and $\theta_3$ on the basis $\eta^8$, $\eta^1$, and $G$ as follows :

\beq
\begin{array}{lll}
{\mathbf O_R}&=& \left(\begin{array}{ccc}
c_1 c_2  & -c_1 s_2 s_3 - s_1 c_3   & -c_1 s_2 c_3 + s_1 s_3     \\
s_1 c_2  & -s_1 s_2 s_3 + c_1 c_3   & -s_1 s_2 c_3 - c_1 s_3     \\
s_2      & c_2 s_3                  & c_2 c_3
\end{array}\right)  \\
& \equiv &
\left(\begin{array}{ccc}
\alpha_8  & \alpha_1  & \alpha_G    \\
\beta_8   & \beta_1   & \beta_G  \\
\gamma_8  & \gamma_1  & \gamma_G
\end{array}\right)~,
\end{array}
\label{mixtfmnmtx}
\eeq
%\end{widetext}
where $c_i\equiv \cos \theta_i,~s_i\equiv \sin \theta_i~(i=1,2,3)$, then
\beq
\left.\begin{array}{rcl}
\eta   &=& \alpha_8 \eta^8 + \alpha_1 \eta^1 + \alpha_G G~, \\
\etap  &=& \beta_8  \eta^8 + \beta_1  \eta^1 + \beta_G G~, \\
\etaz  &=& \gamma_8  \eta^8 +  \gamma_1  \eta^1 + \gamma_G G~.
\end{array}\right.
\label{gzdmixpsmsn}
\eeq
For some purposes, it is convenient to use quark and gluonium basis~\cite{jlRosner1983}
\beq
\left.\begin{array}{rcl}
\eta   &=& \xet N + \yet S + \zet G~, \\
\etap  &=& \xetp N + \yetp S + \zetp G~, \\
\etaz  &=& \xetz N + \yetz S + \zetz G~.
\end{array}\right.
\label{gzdmixofxyz}
\eeq
Here the basis states are
\beq
\left.\begin{array}{rcl}
|N \rangle &\equiv & \frac{1}{\sqrt{2}} |u \overline{u} + d \overline{d}\rangle ~, \\
|S \rangle &\equiv & |s \overline{s}\rangle ~, \\
|G \rangle &\equiv & |\mbox{gluonium}\rangle ~.
\end{array}\right.
\label{basissts}
\eeq

These are related to $\xet$, etc., as follows :
\beq
\left.\begin{array}{rcl}
X_{\xi} &=& \zeta_8 /\sqrt{3} + \sqrt{\frac{2}{3}} \zeta_1~, \\
Y_{\xi} &=&-\sqrt{\frac{2}{3}} \zeta_8  + \zeta_1/\sqrt{3}~, \\
Z_{\xi}  &=& \zeta_G~,
\end{array}\right.
\label{xyzandabc}
\eeq
where $\xi=\eta,~ \etap,~ \etaz,$ and correspondingly $\zeta=\alpha,~ \beta,~\gamma.$ Then following the logic of Ref.~\cite{moxh2024}, constructed are two nonets ${\mathbf V_N}$ and ${\mathbf P_N}$, which can be treated in a way that is akin to octets, then the effective Hamiltonian in Eq.~\eref{effhmtvptype}, can be used formally to acquire the corresponding parametrization for $V P$ final states, the results are summarized in Table~\ref{vpmsnfmmix}.

%\subsection{Three-body decay and relevant study}
Now we discuss the parametrization for charmonium three-body decay.
According to the $C$ parities of the three octets $O_i~(i=1,2,3)$, one of two possible interactions must be adopted~\cite{Haber}:
\begin{widetext}
\beq
\left.\begin{array}{rcl}
\Heff &= & g_{A} \mbox{Tr} [O_1 (O_2 O_3-O_3 O_2)] + g_{A_1} \mbox{Tr}  [O_4 (O_2 O_3 O_1 - O_1 O_3 O_2)]\\
      &  & + g_{A_2} \mbox{Tr}  [O_4 (O_3 O_2 O_1 - O_1 O_2 O_3)]
           + g_{A_3} \mbox{Tr}  [O_4 (O_3 O_1 O_2 - O_2 O_1 O_3)]~,
\end{array}\right.
\label{srbdyasyhmtn}
\eeq
\beq
\left.\begin{array}{rcl}
\Heff &= & g_{S} \mbox{Tr} [O_1 (O_2 O_3+O_3 O_2)] + g_{S_1} \mbox{Tr}  [O_4 (O_2 O_3 O_1 + O_1 O_3 O_2)]\\
      &  & + g_{S_2} \mbox{Tr}  [O_4 (O_3 O_2 O_1 + O_1 O_2 O_3)]
           + g_{S_3} \mbox{Tr}  [O_4 (O_3 O_1 O_2 + O_2 O_1 O_3)] \\
      &  & + g_{S_4} \mbox{Tr} (O_1 O_2) \mbox{Tr}(O_3 O_4)
           + g_{S_5} \mbox{Tr} (O_1 O_3) \mbox{Tr}(O_2 O_4)
           + g_{S_6} \mbox{Tr} (O_1 O_4) \mbox{Tr}(O_2 O_3)~.
\end{array}\right.
\label{srbdysyshmtn}
\eeq
\end{widetext}

For the traceless matrix the last term of Eq.~\eref{srbdysyshmtn} can be eliminated due to Cayley-Burgoyne's identity~\cite{sColeman}
\beq
\sum\limits_{\mbox{\scriptsize six perms}}  \mbox{Tr} [O_i O_j O_k O_l ]
=\sum\limits_{\mbox{\scriptsize three perms}} \mbox{Tr} (O_i O_j) \mbox{Tr} (O_k O_l)~,
\label{mtxidty}
\eeq
where the sum runs over all distinct cyclic permutations of $i,j,k,l=1,2,3,4$.
Adding interactions involving $SU(3)$ singlets is straightforward. To invoke nonet symmetry, simply replace the octets above with nonets. The desired $SU(3)$ breaking can be simulated by taking $O_4$ to be the spurion octet. As an example, the effective Hamiltonian for three pseudoscalar mesons can be expressed as follows :
\beq
\left.\begin{array}{l}
\Heff^{PPP} =g_A \epsilon_{\nu \alpha \beta} \mbox{Tr} \partial^{\nu}N_1
  [\partial^{\alpha} N_2 \partial^{\beta} N_3 - \partial^{\beta} N_3 \partial^{\alpha} N_2]  \\
   +\sum\limits_{B} g_B \epsilon_{\nu \alpha \beta}\mbox{Tr}[S_B (\partial^{\nu} N_2 \partial^{\alpha} N_3 \partial^{\beta} N_1-
                                         \partial^{\beta} N_1 \partial^{\alpha} N_3 \partial^{\nu} N_2 )]~,
\end{array}\right.
\label{hmtnforppp}
\eeq
where the sum runs over all three breaking terms, that is $B=m,e$, and $\mu$; $N_1=N_2=N_3={\mathbf P_N}$. The corresponding parametrization form is summarized in Table~\ref{pppmsnform}.
\begin{table}[hbt]
\caption{\label{pppmsnform} Amplitude parametrization forms for decays of a resonance into three pseudoscalar mesons. Symbols $A$, $D_m$, $D_e$,and $D_{\mu}$, are introduced for simplifying the expression, the relation of them with effective coupling constants $g_A$, $g_m$, $g_e$, and $g_{\mu}$ are indicated in this table. } \center
%The shorthand symbols are defined as $s_{P}=\sin \theta_{P}$, $c_{P}=\cos \theta_{P}$, $s^{-}_{\gamma P}=\sin ( \theta_{\gamma}- \theta_{P})$,$c^{-}_{\gamma P}=\cos ( \theta_{\gamma}- \theta_{P})$, $\sin \theta_{\gamma}\equiv\sqrt{1/3}$, and $\cos \theta_{\gamma}\equiv\sqrt{2/3}$.} \center
\begin{tabular}{lrrrr}\hline \hline
  Final        & \multicolumn{4}{c}{Coupling constant} \\ \cline{2-5}
  state        &$A$            &$D_m$     &$D_e$       &$D_{\mu}$     \\  \cline{2-5}
               &$6\sqrt{2}g_A$ &$6\sqrt{2}g_m$&$6/\sqrt{2} g_e$&$6/\sqrt{2} g_{\mu}$ \\ \hline
$\piz\pip\pim$ &$1$            & $1$   & $1$     & $1$     \\
$\piz\kap\kam$ &$\frac{1}{2}$  & $0$   & $1$     & $2$     \\
$\piz\kaz\kazb$&$-\frac{1}{2}$ & $0$   & $1$     & $0$     \\
$\pip\kam\kaz$ &$-\frac{1}{\sqrt{2}}$  &$0$&$0$ & $-\sqrt{2}$    \\
$\pim\kap\kazb$&$\frac{1}{\sqrt{2}}$   &$0$&$0$ & $\sqrt{2}$    \\
$\eta\kap\kam$ &$\frac{\sqrt{3}}{2}c_P$&$\frac{1}{\sqrt{2}}c^{-}_{\gamma P}$
                              & $s^{-}_{\gamma P}$     & $2\sqrt{3} c_P$     \\
$\etap\kap\kam$&$\frac{\sqrt{3}}{2}s_P$&$-\frac{1}{\sqrt{2}}s^{-}_{\gamma P}$
                              & $c^{-}_{\gamma P}$     & $2\sqrt{3} s_P$     \\
$\eta\kaz\kazb$&$\frac{\sqrt{3}}{2}c_P$&$\frac{1}{\sqrt{2}}c^{-}_{\gamma P}$
                              & $-\sqrt{3} c_P$     & $\sqrt{2} c^{-}_{\gamma P}$     \\
$\etap\kaz\kazb$&$\frac{\sqrt{3}}{2}s_P$&$-\frac{1}{\sqrt{2}}s^{-}_{\gamma P}$
                              & $-\sqrt{3} s_P$     & $-\sqrt{2} s^{-}_{\gamma P}$     \\
$\eta\pip\pim$  & $0$ &$0$ & $s^{-}_{\gamma P}$ & $s^{-}_{\gamma P}$    \\
$\etap\pip\pim$ & $0$ &$0$ & $c^{-}_{\gamma P}$ & $c^{-}_{\gamma P}$    \\
\hline \hline
\end{tabular}
\end{table}
%\subsection{Radiative decay}
At last, we consider the radiative decay. Here the material matter is to express the photon field in the elementary representation of $SU(3)$ group. One choice is define the photon field as $A_{\mu}=\gamma_{\mu} Q$, where $\gamma_{\mu}$ denote the photon field, and $Q$ is the charge matrix in the elementary representation of $SU(3)$ group. As an example, the effective Hamiltonian for $\psi \to \gamma P$ can be expressed as follows :
\beq
\Heff^{\gamma P} =g_0 \cdot \gamma \cdot \mbox{Tr} \{ Q,P\}
     +\sum\limits_{B} g_B \cdot \gamma \cdot \mbox{Tr}[S_B \{ Q,P\}]~,
\label{hmtnforgmap}
\eeq
where $B=m,e$, and $\mu$ as in Eq.~\eref{hmtnforppp}. The corresponding parametrization form is summarized in Table~\ref{gmapdcyform}.
\begin{table}[hbt]
\caption{\label{gmapdcyform} Amplitude parametrization forms for radiative decays of a resonance into $\gamma$ plus a pseudoscalar mesons. Symbols $A$, $D_m$, $D_e$, and $D_{\mu}$, are introduced for simplifying the expression, the relation of them with effective coupling constants $g_0$, $g_m$, $g_e$, and $g_{\mu}$ are indicated in this table. The meanings of $X_{\xi}$, $Y_{\xi}$, and $Z_{\xi}$ ($\xi=\eta,~ \etap,~ \etaz,$) are given in Eq.~\eref{xyzandabc}.} \center
{\footnotesize
\begin{tabular}{lrrrr}\hline \hline
  Final        & \multicolumn{4}{c}{Coupling constant} \\ \cline{2-5}
  state        &$A$            &$D_m$     &$D_e$       &$D_{\mu}$     \\  \cline{2-5}
               &$\sqrt{2}g_0/3$&$4 g_m/3$   &$2 g_e/3$     &$4 g_{\mu}/3$ \\ \hline
$\gamma \piz$  &$3 $           & $\frac{3}{2\sqrt{2}}$ & $\frac{3}{\sqrt{2}}$ & $\frac{3}{2\sqrt{2}}$    \\
$\gamma \eta$  &$\xet-\sqrt{2}\yet$  &$\frac{1}{2\sqrt{2}}\xet+\yet$
 &$\frac{5}{\sqrt{2}}\xet+\yet$   &$\frac{5}{2\sqrt{2}}\xet-\yet$   \\
$\gamma \etap$ &$\xetp-\sqrt{2}\yetp$  &$\frac{1}{2\sqrt{2}}\xetp+\yetp$
 &$\frac{5}{\sqrt{2}}\xetp+\yetp$   &$\frac{5}{2\sqrt{2}}\xetp-\yetp$    \\
$\gamma \eta_0$&$\xetz-\sqrt{2}\yetz$  &$\frac{1}{2\sqrt{2}}\xetz+\yetz$
 &$\frac{5}{\sqrt{2}}\xetz+\yetz$   &$\frac{5}{2\sqrt{2}}\xetz-\yetz$   \\
\hline \hline
\end{tabular} }
\end{table}
%\subsection{Comment}
In this Letter, we have established a systematic parametrization scheme for all charmonium hadronic decays. The scheme is simple and clear, as long as the breaking effect form is fixed, with expression of effective Hamiltonian, no tour de force is needed, the simple algebras yields the amplitude parametrization. Moreover,
in principle any $(n+1)$-body Hamiltonian can be used to obtain second-order correction for $n$-body symmetry breaking effect or $n$-order correction for two-body symmetry breaking effect. The prominent merit of the method lies in that the feature due to symmetry has been taken into account, what need to be considered furthermore are solely the dynamics aspect of charmonium decay. Although the addition of unknown parameters make the experimental application marginal or even impractical, such a systematic parametrization method blaze a new avenue to describe the all kinds of charmonium decay. Theoretically, it stimulates us to develop a method to figure out the relation of these parameters, and at the same time provides a new angle to view decay mechanism.

This work is supported in part by National Key Research and Development Program of China under Contracts No.~2023YFA1606003 and No.~2023YFA1606000.

\end{document}